\documentclass[useAMS,usenatbib,usegraphicx]{mn2e}
\usepackage{journal}
\usepackage{amsmath}
\usepackage{amssymb}
\renewcommand{\vec}{\boldsymbol}
\newcommand{\bm}{\vec{\overline{B}}}
\newcommand{\jm}{\vec{\overline{j}}}
\newcommand{\Em}{\vec{\overline{E}}}
\newcommand{\emf}{\vec{\mathcal{E}}}

\title[Rotational threshold in global numerical dynamo simulations]{Rotational 
threshold in global numerical dynamo simulations}
\author[M. Schrinner]{M. Schrinner$^{1}$\thanks{E-mail:
martin@schrinner.eu}\\
$^{1}$MAG(ENS/IPGP),LRA, \'Ecole Normale Sup\'erieure, 24 Rue Lhomond, 
  75252 Paris Cedex 05, France}

\begin{document}
\date{Accepted . Received; in original form}
\pagerange{\pageref{firstpage}--\pageref{lastpage}} \pubyear{2012}
\maketitle
\label{firstpage}
\begin{abstract}
Magnetic field observations of low-mass stars reveal an increase of magnetic 
activity with increasing rotation rate. The so-called activity-rotation 
relation is usually attributed to changes in the underlying dynamo
processes generating the magnetic field. We examine the dependence of the 
field strength on rotation in global numerical dynamo models and interpret 
our results on the basis of energy considerations. In agreement with the 
scaling law proposed by \cite{christensen06}, the field strength in 
our simulations is set by the fraction of the available 
power used for the magnetic field generation. This is controlled by the dynamo 
efficiency calculated as the ratio of Ohmic to total dissipation in 
our models. The dynamo efficiency grows strongly with increasing rotation rate 
at a Rossby number of 0.1 until it reaches its upper bound of one and becomes 
independent of rotation. This gain in efficiency is related to the strong 
rotational dependence of the mean electromotive force in this parameter 
regime. For multipolar models at Rossby numbers clearly larger than 
0.1, on the other hand, we do not find a systematic dependence of the field 
strength on rotation. Whether the enhancement of the dynamo efficiency found
in our dipolar models explains the observed activity-rotation 
relation needs to be further assessed.
      
\end{abstract}

\section{Introduction}
There is considerable observational evidence that magnetic activity on low 
mass stars increases with increasing stellar rotation rate until it saturates 
and reaches a constant level for very fast rotating 
stars \citep[e.g.][]{reiners12a}. This so-called activity-rotation relation 
is primarily based on observations of chromospheric or coronal magnetic 
activity indicators, i.e. on chromospheric or coronal 
emission \citep{skumanich72,noyes84,delfosse98,pizzolato03}. Furthermore, it 
is supported by magnetic flux measurements in stars of spectral 
types G-M \citep{saar01,reiners09a}. \cite{noyes84} pointed out that 
chromospheric emission and thus magnetic activity depending on 
spectral type and rotation is well described by a single parameter, the 
Rossby number \(Ro=P_\mathrm{rot}/\tau_c\), with \(P_\mathrm{rot}\) being the 
observed rotation period of a star and \(\tau_c\) a convective turnover time 
derived from mixing-length theory. The observed increase of magnetic activity
with decreasing Rossby number is usually attributed to changes in the
underlying dynamo processes generating the magnetic field
\citep{donati09,reiners12a}. However, until now, theoretical arguments 
explaining the activity-rotation relation are poorly developed and formulated
only on a heuristic level. 

The standard argument worked out in the framework 
of linear mean-field theory refers to the so-called dynamo number, \(D\), a 
dimensionless parameter which determines the distance to the dynamo threshold 
of a given dynamo model. If \(D\) exceeds some critical value, \(D>D_c\), 
a small initial magnetic perturbation may grow exponentially, i.e. the 
field-free state is linearly unstable. The larger \(D\), the larger are the 
growth rates of the magnetic field expected in this scenario. Under some 
simplifying assumptions, \(D\) scales inversely proportional to the square of 
the Rossby number \citep{noyes84b}. Therefore, one might argue that dynamo 
action can be easier excited and thus leads to larger field strengths if the 
Rossby number decreases. 

The principal objection against this argument is that it is entirely based on
linear, kinematic theory. For stellar dynamos with \(D\gg D_c\), the magnetic 
field would grow rapidly until the Lorentz force changes the 
velocity field and as a result the magnetic field saturates. In this 
dynamical regime, linear theory is in general no longer applicable.  Even if 
the dynamo number predicted the onset of dynamo action properly, its 
predictive power for the saturation level of the magnetic field would 
remain uncertain. 
    
A successful scaling law for the field strength of fast rotating stars and 
planets, on the other hand, was presented by \cite{christensen09}. They 
showed that a scaling originally derived from geodynamo models in the 
Boussinesq approximation \citep{christensen06} also applies to certain classes 
of stars. The scaling law reads
\begin{equation}
B^2/(2\mu_0)\,\sim\,f_\mathrm{ohm}\varrho^{(1/3)}(q_c\,L/H_T)^{(2/3)}
\label{eq:2}
\end{equation}
and is based on the available energy flux \(q_c\), which sets the field 
strength apart from the density, \(\varrho\), and a convective
length scale relative to the temperature scale height, \(L/H_T\). The 
coefficient \(f_\mathrm{ohm}\le1\) gives the ratio of Ohmic to total 
dissipation and was set to unity in this context. \cite{christensen09}
highlighted two remarkable findings: First, the flux based scaling law is at 
first glance independent of rotation, and second, it is only applicable to 
fast rotating stars and planets.   

In the following we will argue that the flux based scaling law (\ref{eq:2})
is in fact valid for a wide range of rotation rates, but it is not 
independent of rotation. On the contrary, we claim that the rotational 
dependence is captured by \(f_\mathrm{ohm}\) and simply eliminated by setting 
\(f_\mathrm{ohm}\) to unity. This indeed seems to be justified only for 
stars in the rotationally saturated regime. Our analysis is based on 30 
numerical dynamo models in the Boussinesq approximation. The modeling
strategy and the models are briefly described in the next section. Hereafter, 
we present results for \(f_\mathrm{ohm}\) revealing its dependence on rotation 
rate and discuss the implications of our finding for the activity-rotation 
relation.  
  
\section{Dynamo Calculations}
Our dynamo models are solutions of the MHD-equations for a conducting 
Boussinesq fluid in a rotating spherical shell. Convection is driven by an 
applied temperature difference \(\Delta T\) between the inner boundary at 
radius \(r_i\) and the outer boundary at \(r_o\). 
The governing equations for the velocity \(\vec{v}\), the 
magnetic field \(\vec{B}\), and the temperature \(T\) written in a 
dimensionless form proposed by \cite{olson99} are
\begin{eqnarray}
E\left(\frac{\partial\vec{v}}{\partial t}+\vec{v}\cdot\nabla\vec{v}-\nabla^2\vec{v}\right)
+2\vec{z}\times\vec{v}+\nabla P =\nonumber\\
Ra\frac{\boldsymbol{r}}{r_o}T
+\frac{1}{Pm}(\nabla\times\vec{B})\times\vec{B}\, ,\label{eq:4}\\
\frac{\partial T}{\partial t}+\vec{v}\cdot\nabla T  =
\frac{1}{Pr}\nabla^2 T \, ,\label{eq:6}\\
\frac{\partial\vec{B}}{\partial t}= \nabla\times(\vec{v}\times\vec{B})
+\frac{1}{Pm}\nabla^2\vec{B},\label{eq:8}\\
\nabla\cdot\vec{v}=0,\hspace{0.5cm}\nabla\cdot\vec{B}=0.\label{eq:10}
\end{eqnarray}  
They are controlled by four dimensionless parameters, the Ekman number 
\(E=\nu/\Omega L^2\), the (modified) Rayleigh number 
\(Ra=\alpha_T g_o\Delta T L/\nu\Omega\), the Prandtl number \(Pr=\nu/\kappa\), 
and the magnetic Prandtl number \(Pm=\nu/\eta\). In these definitions, \(L\) 
denotes the width of the convection zone, \(\Omega\) stands for the angular 
velocity, \(\alpha\) is the thermal expansion coefficient, \(g_o\) is the 
gravitational acceleration at the outer boundary, and \(\nu\), \(\eta\), 
\(\kappa\) are the kinematic viscosity, the magnetic and the 
thermal diffusivity.  The mechanical boundary conditions are no slip and 
the magnetic field continues as a potential field outside the fluid shell. 

In the following paragraph we need to define some further quantities which are 
used throughout in the paper. The time-averaged ratio of ohmic to total 
dissipation, \(f_\mathrm{ohm}\), is computed as 
\begin{equation}
f_\mathrm{ohm}=\frac{W_J}{W_b}
\label{eq:14a}
\end{equation}
with the rate of ohmic dissipation
\begin{equation}
W_J=\frac{1}{Pm^2\,E}\int\,(\nabla\times\,B)^2\,\mathrm{dv},
\label{eq:14b}
\end{equation} 
and the power \(W_b\) generated by buoyancy forces,
\begin{equation}
W_b=\frac{Ra}{E}\int\,\frac{r}{r_o}\,v_r\,T\,\mathrm{dv}.
\label{eq:14c}
\end{equation}
 
Moreover, we define a Rossby number for our models similar to the observational
one. It is given by the ratio of the Rossby radius to a typical 
convective length scale \(\ell_c\), \(Ro_\ell=v_\mathrm{rms}/(\Omega\ell_c)\),
where \(v_\mathrm{rms}\) stands for the rms velocity of the flow and 
\(\ell_c\) is derived from the kinetic energy spectrum 
\citep{schrinner12,christensen06}.  A non-dimensional measure for the 
convective energy flux is the Nusselt number, \(Nu\),  defined as the ratio 
of the total heat flow to the conducted heat flow. Finally,  we note that 
the temperature scale height \(H_T\) for our models is given 
by \(H_T=c_P/(\alpha g_o)\) with the heat capacity \(c_P\). 

We aim at comparing models which differ only in their rotation rates. 
Therefore, we keep the thermodynamically available energy flux and the 
diffusivities for a sequence of models constant and vary only the angular 
velocity. Translated in non-dimensional quantities, this means: We keep (in a 
first approximation) the Rayleigh number over some critical Rayleigh number
\(Ra_c\), the Prandtl number and the magnetic Prandtl number constant and 
change successively the Ekman number.  More precisely, we try to keep the  
Nusselt number \(Nu\) for a sequence of models constant. Because \(Nu\) is an 
output parameter in our simulations and only roughly determined by \(Ra\) 
normalized by its critical value, \(Ra/Ra_c\) has to be adjusted accordingly.
Some more details are given in the Appendix.
 
Due to computational limitations all current numerical dynamo simulations run 
in a parameter regime which is not appropriate for stellar interiors.  
Moreover, Boussinesq models do not account for the strong density variation in
stars and thus certainly do not reproduce stellar dynamo processes in  
realistic detail. However, our models are adequate to study the flux-based 
scaling law which was originally derived from Boussinesq models, too.   

\section{Results}
We considered sequences of models with \(Nu\approx 2.2\), \(3.5\), and \(7\)
and \(Pm\) varying between \(3\) and \(7\). The Prandtl 
number \(Pr\) was always set to unity in our simulations. For given \(Pm\) 
and \(Nu\), we obtained a \textit{sequence of models} by varying the 
Ekman number between \(E=10^{-3}\) and \(E=10^{-5}\); some models with 
\(E=3\cdot 10^{-3}\) and \(E=3\cdot 10^{-6}\) could also be added to our 
sample. On the other hand, simulations with \(Nu\approx 7\) and 
\(E\le 3\times 10^{-5}\) were numerically not feasible. The magnetic Reynolds
number of our models, \(Rm=v_\mathrm{rms}\,L/\eta\),  is always larger than 
100 and thus far above the minimum value of \(Rm=40\) needed to obtain dynamo
action in this setting \citep{olson06}. We show in Appendix A that 
there is a slight but systematic increase of \(Rm\) with rotation rate for 
a sequence of models at constant Nusselt and Prandtl numbers.  

\begin{figure}
\includegraphics[scale=0.7]{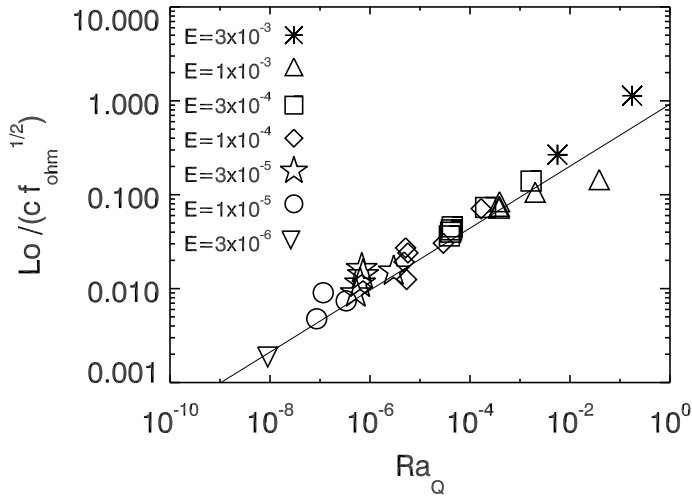}
\caption{\(Lo/(c\,f_\mathrm{ohm}^{1/2})\) versus the flux based 
Rayleigh number. The constant \(c\) was chosen to be 0.92 for dipolar 
and 0.48 for multipolar models \citep{christensen06,christensen10}. The solid 
line is not a fit but the prediction of equation (\ref{eq:14}).}
\label{fig1}
\end{figure}

Following \cite{christensen06} a non-dimensional form of the flux-based 
scaling law is obtained by dividing relation (\ref{eq:2}) 
by \(\varrho\,\Omega^2 L^2\),
\begin{equation}
B^2/(2\mu_0\varrho\Omega^2 L^2)\,\sim\,f_\mathrm{ohm}(q_c/(\varrho\Omega^3L^3)\,L/H_T)^{(2/3)}.
\label{eq:12}
\end{equation}
With the Lorentz number \(Lo=B/(\mu\varrho\Omega^2L^2)^{1/2}\), 
the flux based Rayleigh number 
\(Ra_Q=q_c\alpha g_o/(4\pi\varrho c_P\Omega^3 L^2)\), and 
\(L/H_T=L\,\alpha g_o/c_P\), relation (\ref{eq:12}) may simply be written as
\begin{equation}
Lo/f_\mathrm{ohm}^{1/2}=c\,Ra_Q^{1/3}.
\label{eq:14}
\end{equation} 
\cite{christensen06} found the prefactor \(c\) to be 0.92 for models with a
predominantly dipolar field geometry and a slightly lower value of \(c=0.48\) 
was given by \cite{christensen10} for multipolar models. In Fig. \ref{fig1} we 
plotted \(Lo/(c\,f_\mathrm{ohm}^{1/2})\) against \(Ra_Q\) in logarithmic 
scales for our sample of models. Independent of their Ekman number, 
all models are in agreement with the scaling (\ref{eq:14}) 
indicated in Fig. \ref{fig1} by the solid line. Apparently, the flux-based 
scaling law holds for all models independently of their rotation rates. 

\begin{figure}
\includegraphics{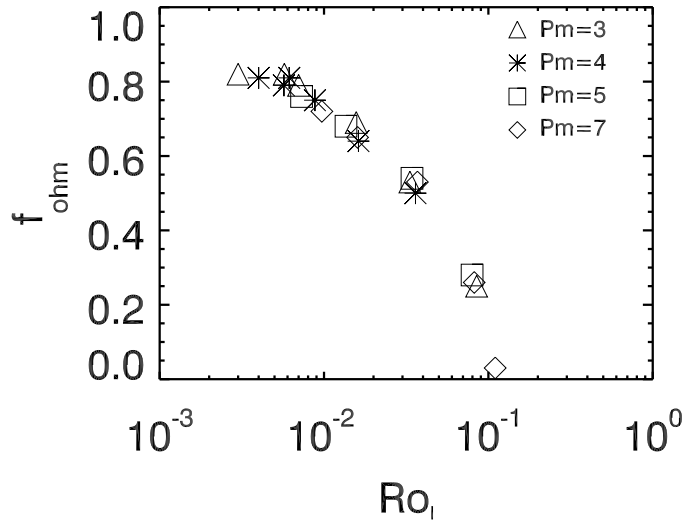}
\caption{The ratio of ohmic to total dissipation, \(f_\mathrm{ohm}\), versus 
the Rossby number for sequences of models with \(Nu=2.2\) and various \(Pm\). }
\label{fig2}
\end{figure}

However, the ratio of ohmic to total dissipation, \(f_\mathrm{ohm}\), may 
increase drastically with rotation rate, as demonstrated in 
Fig. \ref{fig2}. Shown is \(f_\mathrm{ohm}\) versus the Rossby number 
for sequences of models with \(Nu=2.2\) and various \(Pm\). At 
\(Ro_\ell\approx 0.12\), the rate of ohmic diffusion increases rapidly 
until the steep slope flattens and \(f_\mathrm{ohm}\) saturates for 
lower Rossby numbers. On the other hand, a strong dependence of 
\(f_\mathrm{ohm}\) on \(Pm\) cannot be inferred from Fig. \ref{fig2}. 

\begin{figure}
\includegraphics{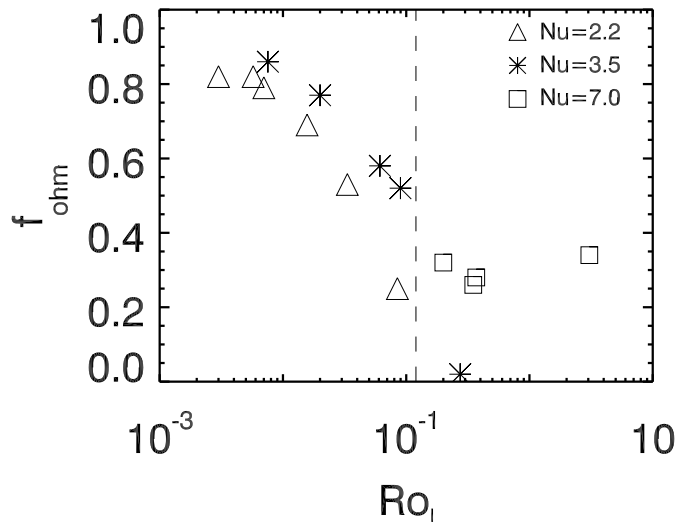}
\caption{The dynamo efficiency \(f_\mathrm{ohm}\) versus the Rossby 
number for sequences of models with \(Pm=3\) and Nusselt numbers \(Nu=2.2\), 
\(Nu=3.5\), and \(Nu=7\). The vertical dashed line indicates the 
boundary between the dipolar (\(Ro_\ell<0.12\)) and multipolar dynamo 
regime.}
\label{fig3}
\end{figure}

The strong increase of \(f_\mathrm{ohm}\) at \(Ro_\ell\lesssim 0.12\) 
coincides with a transition from multipolar dynamo models at higher Rossby 
number to models with a dipole dominated magnetic field. In fact, the 
dependence of \(f_\mathrm{ohm}\) on the rotation rate changes crucially 
at the regime boundary as shown in Fig. \ref{fig3}. For a sequence of 
predominantly dipolar models with a slightly higher Nusselt 
number, \(Nu=3.5\), the increase of \(f_\mathrm{ohm}\) is qualitatively 
reproduced. However, the amplified convective energy flux shifts 
the sequence towards higher Rossby numbers. For an even higher Nusselt 
number, \(Nu=7\), the sequence of models falls entirely in the 
multipolar regime and a systematic increase of \(f_\mathrm{ohm}\) with 
rotation rate is no longer observed.

\section{Discussion and conclusions}
In an equilibrium state, the energy released by buoyancy in our models is 
dissipated by viscous dissipation and ohmic diffusion. The latter requires
that a magnetic field is built up by dynamo action and the rate of ohmic 
dissipation determines the fraction of the available power used for 
the magnetic field generation. For fast rotators \(f_\mathrm{ohm}\) increases,
this means that a larger fraction of the available power is converted to 
magnetic energy and dynamo action becomes more efficient. According to 
relation (\ref{eq:2}), the growth of \(f_\mathrm{ohm}\) visible in 
Fig. \ref{fig2} leads to an increase of the average magnetic field strength 
by an order of magnitude. Because \(f_\mathrm{ohm}\) is bound by one, the 
field strength saturates for even higher rotation rates and then becomes 
independent of rotation. Both, the increase  at \(Ro_\ell\approx 0.1\) and 
the saturation of the field strength are in good agreement with observations. 
Yet, not much is known about the saturation level of the magnetic field in 
slowly rotating stars, except that it falls below the value 
predicted by (\ref{eq:2}) with \(f_\mathrm{ohm}=1\) \citep{christensen09}. 
With a Rossby number of \(Ro_\ell > 0.5\) \citep{reiners12}, the Sun is an 
example for a slow rotator. Assuming an energy flux 
of \(q_c=63\,\mathrm{MW/m^2}\), a mean density 
of \(\varrho=1.4\,\mathrm{kg}/\mathrm{m}^3\) and an average internal 
field of \(B=0.063\, \mathrm{T}\) \citep{christensen09}, relation 
(\ref{eq:2})  requires \(f_\mathrm{ohm}\approx 0.07\). This would be
consistent with the range of \(f_\mathrm{ohm}\) presented in this study. We 
note, however, that the flux based scaling law is somewhat at odds with the 
estimate of \(f_\mathrm{ohm}=10^{-3}\) derived by \cite{rempel06} 
from dynamic flux-transport solar dynamo models. 

The decline of \(f_\mathrm{ohm}\) at \(Ro_\ell\approx 0.1\) visible 
in Fig. \ref{fig2} and Fig. \ref{fig3} is related to a rotational dynamo 
threshold. It is characterized by a minimum magnetic Prandtl number, 
\(Pm_\mathrm{crit}\), below which self-sustained dynamo action does 
not occur. \cite{COG99} found that \(Pm_\mathrm{crit}\) is 
a function of only the Ekman number and varies in the dipolar regime as 
\begin{equation}
Pm_\mathrm{crit}=450\,E^{0.75}.
\label{eq:16}
\end{equation}
Models with given diffusivities approach this dynamo 
threshold if their rotation rate is decreased and \(f_\mathrm{ohm}\) drops 
to zero.

Equation (\ref{eq:16}) is independent of any velocity amplitude and 
does not relate the rotational threshold to a given Rossby number. Hence, 
particular low values for \(f_\mathrm{ohm}\) could in principle be found at 
any Rossby number (and at any \(Rm\) ). However, we could not confirm 
equation (\ref{eq:16}) for the multipolar dynamo regime 
where \(f_\mathrm{ohm}\) remains low and does not change significantly with 
rotation rate (see Fig. \ref{fig3}).
Attempts to identify a similar decline of \(f_\mathrm{ohm}\) close to a 
rotational threshold at much lower Rossby numbers also failed. Models in this 
parameter regime with a lower Ekman and magnetic Prandtl often bifurcate 
subcritically \citep{morinv10}. Consequently, the magnetic field strength and 
\(f_\mathrm{ohm}\) remain high and collapse abruptly to zero only beyond
the dynamo threshold.  

What explains the rotational threshold and the high rotational sensitivity 
of \(f_\mathrm{ohm}\) for models in the dipolar regime close 
to \(Ro_\ell=0.1\)? Dipolar models typically exhibit different 
characteristic length scales for their velocity and their magnetic field 
with \(\ell_V<\ell_B\). The balance of the advection and the diffusion term 
in the induction equation then leads to a modified magnetic Reynolds number, 
\(\widetilde{Rm}=v_\mathrm{rms}\,\ell_B^2/(\eta\,\ell_V)\), which needs to 
exceed a critical value for the onset of dynamo action. For given 
diffusivities, \(\ell_B^2\) is inversely proportional to \(v_\mathrm{rms}\) 
\citep{christensen04}. Therefore, an increase of the velocity amplitude 
is compensated by smaller \(\ell_B\) and does not change \(\widetilde{Rm}\). 
This heuristic argument might explain why equation (\ref{eq:16}) is independent
of any velocity amplitude and holds only in the dipolar regime. The strong
dependence of the dynamo efficiency on rotation rate, however, requires some 
further explanation. 

\begin{figure}
\includegraphics[scale=0.6]{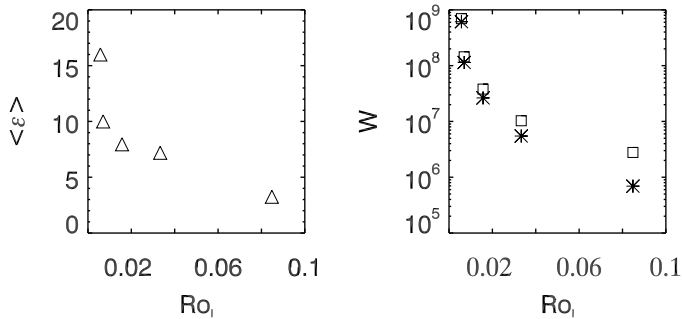}
\caption{Left: Time-averaged rms values of the electromotive force in units of
\([\nu/L\,(\varrho\eta\mu\Omega)^{1/2}]\) for a sequence of models with 
\(Nu\approx 2.2\) and \(Pm=4\) versus the Rossby number. Right: The rate of 
Ohmic dissipation (stars) and total dissipation (squares) in units 
of \(\varrho\nu^3/L\) versus \(Ro_\ell\) for the same sequence of models.}
\label{fig4}
\end{figure}

Dynamo models in the dipolar regime with \(Ro_\ell\lesssim 0.1\) 
may be adequately described in the framework of mean-field theory 
\citep{raedler80,schrinner07,schrinner11b,schrinner11c}. The mean-field 
formalism provides useful concepts to better understand the influence of 
rotation on the dynamo processes in our models. It is usually set up by 
splitting the velocity and the magnetic field in a mean and a residual 
component varying on different length scales, 
\(\vec{v}=\vec{\overline{V}}+\vec{v'}\) and  
\(\vec{B}=\vec{\overline{B}}+\vec{b'}\).  Mean quantities denoted here by 
an overbar may be thought of as combined azimuthal and time averages 
\citep{schrinner11b}. They vary on a scale similar to the system size 
\(L\). Fluctuating quantities, on the other hand, vary on the scale of the 
velocity field, \(\ell_V\). We emphasize that this is not an assumption but 
a result obtained from direct numerical simulations in a particular parameter 
regime. The mean flow \(\vec{\overline{V}}\) is 
negligible for the models considered here \citep{olson99, schrinner07}. 
Therefore, the induction equation (\ref{eq:4}) separated for the mean and the 
residual component may be written as
\begin{eqnarray}
\frac{\partial\bm}{\partial t}-\frac{1}{Pm}\nabla^2\bm & = & \nabla\times\overline{\vec{v}\times\vec{b'}}\\
\label{eq:18}
\wedge\,\frac{\partial\vec{b'}}{\partial t}-\nabla\times(\vec{v}\times\vec{b'})'-\frac{1}{Pm}\nabla^2\vec{b'} & = & \nabla\times(\vec{v}\times\bm). 
\label{eq:20}
\end{eqnarray}
The residual magnetic field \(\vec{b'}\) is diffusive on the length scale 
\(\ell_V\) and would rapidly decay without the source term on the right-hand 
side of (\ref{eq:20}). Thus, the magnetic field-generation depends decisively
on \(\bm\) and in particular on the so-called mean electromotive force, 
\(\emf=\overline{\vec{v}\times\vec{b'}}\). Similarly, the magnetic 
energy density is dominated by the mean field. The energy equation 
for \(\bm\) reads
\begin{equation}
\frac{d}{dt}\int_\infty\frac{\bm^2}{2}\,\mathrm{d}v=-\int_\mathcal{V}\jm\cdot\Em\,\mathrm{d}v,
\label{eq:22}
\end{equation}
where \(\Em\) is the mean electrical field, \(\jm\) is the mean current 
density and \(\mathcal{V}\) denotes the volume of the fluid shell.
With Ohm's law, 
\begin{equation}
\jm=Pm\,(\Em+\emf),
\end{equation}
equation (\ref{eq:22}) yields
\begin{equation}
\frac{1}{Pm}\int_\mathcal{V}\jm^2\,\mathrm{d}v=\int_\mathcal{V}\jm\cdot\emf\,\mathrm{d}v.
\end{equation}
for an equilibrium state. Hence, also the mean Ohmic diffusion is controlled
by the electromotive force, i.e. by the correlation of the residual
velocity and the residual magnetic field, and thus linked to rotation. It is
expected that rotation strengthens the correlation between 
\(\vec{v}\) and \(\vec{b'}\). Indeed, \(\emf\) grows with rotation rate for a 
sequence of models with fixed \(Nu\), as shown in Fig. \ref{fig4}. Therefore, 
also \(W_J\) and eventually the dynamo efficiency \(f_\mathrm{ohm}\) 
increase with decreasing Rossby number. We note, however, that also \(W_b\) 
(in units of \(\varrho\nu^3/L\)) increases for this sequence of models, 
though somewhat slower than \(W_J\).

For clarification, we stress that the decrease of the dynamo efficiency
with increasing Rossby number in our models is not caused by an 
\(Rm\)-dependent quenching of the electromotive force, which is sometimes 
called catastrophic quenching \citep[see][and references therein]{branden05}.
In contrast to the catastrophic quenching scenario the mean electromotive 
force increases with \(Rm\) in our simulations (see also Appendix A).

In summary, the field strength of our models is set by the available energy
flux and via \(f_\mathrm{ohm}\) by the rotation rate. The dynamo efficiency 
\(f_\mathrm{ohm}\) increases strongly with rotation rate at 
\(Ro_\ell\approx 0.1\) and saturates at smaller Rossby numbers. The high 
rotational sensitivity of \(f_\mathrm{ohm}\) is related to a rotational dynamo 
threshold and finally to the strong dependence of the mean electromotive force 
on rotation in this parameter regime. For multipolar dynamos at higher Rossby 
number, however, the dynamo efficiency seems to be almost independent of 
rotation. Similarities with the observed activity rotation relation are 
encouraging and need to be further assessed.    

\section*{Acknowledgments}

This study was initiated by an interesting discussion with 
E. Dormy, J. Morin, and J.F. Donati. MS is grateful for financial support 
from the DFG fellowship SCHR 1299/1-1. Computations were performed at CINES 
and CEMAG computing centres. 
\bibliographystyle{mn2e}
\bibliography{schrinner}

\begin{thebibliography}{}

\bibitem[\protect\citeauthoryear{{Brandenburg} \& {Subramanian}}{{Brandenburg}
  \& {Subramanian}}{2005}]{branden05}
{Brandenburg} A.,  {Subramanian} K.,  2005, \physrep, 417, 1

\bibitem[\protect\citeauthoryear{{Busse}}{{Busse}}{1970}]{busse70}
{Busse} F.~H.,  1970, Journal of Fluid Mechanics, 44, 441

\bibitem[\protect\citeauthoryear{{Christensen}, {Olson} \&
  {Glatzmaier}}{{Christensen} et~al.}{1999}]{COG99}
{Christensen} U.,  {Olson} P.,    {Glatzmaier} G.~A.,  1999, Geophysical
  Journal International, 138, 393

\bibitem[\protect\citeauthoryear{{Christensen}}{{Christensen}}{2010}]{christen%
sen10}
{Christensen} U.~R.,  2010, \ssr, 152, 565

\bibitem[\protect\citeauthoryear{{Christensen} \& {Aubert}}{{Christensen} \&
  {Aubert}}{2006}]{christensen06}
{Christensen} U.~R.,  {Aubert} J.,  2006, Geophy. J. Int., 166, 97

\bibitem[\protect\citeauthoryear{{Christensen}, {Holzwarth} \&
  {Reiners}}{{Christensen} et~al.}{2009}]{christensen09}
{Christensen} U.~R.,  {Holzwarth} V.,    {Reiners} A.,  2009, \nat, 457, 167

\bibitem[\protect\citeauthoryear{{Christensen} \& {Tilgner}}{{Christensen} \&
  {Tilgner}}{2004}]{christensen04}
{Christensen} U.~R.,  {Tilgner} A.,  2004, \nat, 429, 169

\bibitem[\protect\citeauthoryear{{Delfosse}, {Forveille}, {Perrier} \&
  {Mayor}}{{Delfosse} et~al.}{1998}]{delfosse98}
{Delfosse} X.,  {Forveille} T.,  {Perrier} C.,    {Mayor} M.,  1998, \aap, 331,
  581

\bibitem[\protect\citeauthoryear{{Donati} \& {Landstreet}}{{Donati} \&
  {Landstreet}}{2009}]{donati09}
{Donati} J.-F.,  {Landstreet} J.~D.,  2009, \araa, 47, 333

\bibitem[\protect\citeauthoryear{{Krause} \& {R{\"a}dler}}{{Krause} \&
  {R{\"a}dler}}{1980}]{raedler80}
{Krause} F.,  {R{\"a}dler} K.,  1980, {Mean-field magnetohydrodynamics and
  dynamo theory}.
Oxford: Pergamon Press

\bibitem[\protect\citeauthoryear{{Morin} \& {Dormy}}{{Morin} \&
  {Dormy}}{2009}]{morinv10}
{Morin} V.,  {Dormy} E.,  2009, International Journal of Modern Physics B, 23,
  5467

\bibitem[\protect\citeauthoryear{{Noyes}, {Hartmann}, {Baliunas}, {Duncan} \&
  {Vaughan}}{{Noyes} et~al.}{1984}]{noyes84}
{Noyes} R.~W.,  {Hartmann} L.~W.,  {Baliunas} S.~L.,  {Duncan} D.~K.,
  {Vaughan} A.~H.,  1984, \apj, 279, 763

\bibitem[\protect\citeauthoryear{{Noyes}, {Weiss} \& {Vaughan}}{{Noyes}
  et~al.}{1984}]{noyes84b}
{Noyes} R.~W.,  {Weiss} N.~O.,    {Vaughan} A.~H.,  1984, \apj, 287, 769

\bibitem[\protect\citeauthoryear{{Olson} \& {Christensen}}{{Olson} \&
  {Christensen}}{2006}]{olson06}
{Olson} P.,  {Christensen} U.~R.,  2006, Earth and Planetary Science Letters,
  250, 561

\bibitem[\protect\citeauthoryear{{Olson}, {Christensen} \&
  {Glatzmaier}}{{Olson} et~al.}{1999}]{olson99}
{Olson} P.,  {Christensen} U.~R.,    {Glatzmaier} G.~A.,  1999, \jgr, 104,
  10383

\bibitem[\protect\citeauthoryear{{Pizzolato}, {Maggio}, {Micela}, {Sciortino}
  \& {Ventura}}{{Pizzolato} et~al.}{2003}]{pizzolato03}
{Pizzolato} N.,  {Maggio} A.,  {Micela} G.,  {Sciortino} S.,    {Ventura} P.,
  2003, \aap, 397, 147

\bibitem[\protect\citeauthoryear{{Reiners}}{{Reiners}}{2012}]{reiners12a}
{Reiners} A.,  2012, Living Reviews in Solar Physics, 9, 1

\bibitem[\protect\citeauthoryear{{Reiners}, {Basri} \& {Browning}}{{Reiners}
  et~al.}{2009}]{reiners09a}
{Reiners} A.,  {Basri} G.,    {Browning} M.,  2009, \apj, 692, 538

\bibitem[\protect\citeauthoryear{{Reiners}, {Joshi} \& {Goldman}}{{Reiners}
  et~al.}{2012}]{reiners12}
{Reiners} A.,  {Joshi} N.,    {Goldman} B.,  2012, \aj, 143, 93

\bibitem[\protect\citeauthoryear{{Rempel}}{{Rempel}}{2006}]{rempel06}
{Rempel} M.,  2006, \apj, 647, 662

\bibitem[\protect\citeauthoryear{{Saar}}{{Saar}}{2001}]{saar01}
{Saar} S.~H.,  2001, in {Garcia Lopez} R.~J.,  {Rebolo} R.,   {Zapaterio
  Osorio} M.~R.,  eds, 11th Cambridge Workshop on Cool Stars, Stellar Systems
  and the Sun Vol.~223 of Astronomical Society of the Pacific Conference
  Series, {Recent Measurements of (and Inferences About) Magnetic Fields on K
  and M Stars (CD-ROM Directory: contribs/saar1)}.
p.~292

\bibitem[\protect\citeauthoryear{{Schrinner}}{{Schrinner}}{2011}]{schrinner11b}
{Schrinner} M.,  2011, \aap, 533, A108+

\bibitem[\protect\citeauthoryear{{Schrinner}, {Petitdemange} \&
  {Dormy}}{{Schrinner} et~al.}{2012}]{schrinner12}
{Schrinner} M.,  {Petitdemange} L.,    {Dormy} E.,  2012, \apj, 752, 121

\bibitem[\protect\citeauthoryear{{Schrinner}, {R{\"a}dler}, {Schmitt},
  {Rheinhardt} \& {Christensen}}{{Schrinner} et~al.}{2007}]{schrinner07}
{Schrinner} M.,  {R{\"a}dler} K.-H.,  {Schmitt} D.,  {Rheinhardt} M.,
  {Christensen} U.~R.,  2007, Geophys. Astrophys. Fluid Dyn., 101, 81

\bibitem[\protect\citeauthoryear{{Schrinner}, {Schmitt} \& {Hoyng}}{{Schrinner}
  et~al.}{2011}]{schrinner11c}
{Schrinner} M.,  {Schmitt} D.,    {Hoyng} P.,  2011, Physics of the Earth and
  Planetary Interiors, 188, 185

\bibitem[\protect\citeauthoryear{{Skumanich}}{{Skumanich}}{1972}]{skumanich72}
{Skumanich} A.,  1972, \apj, 171, 565

\end{thebibliography}
\appendix
\section{Scaling of \(R\lowercase{a}/R\lowercase{a_c}\) and \(R\lowercase{m}\) 
at constant Nusselt number}
We use scaling laws given by \cite{christensen06} to show that the ratio
\(Ra/Ra_c\) decreases slightly with rotation rate for a sequence of models 
at constant Nusselt and Prandtl numbers, whereas the magnetic Reynolds
number increases.  

The Nusselt number scaling proposed by \cite{christensen06} may be written as
\begin{equation}
Nu-1\sim Ra\,E,
\label{ap:2}
\end{equation}
provided that \(Nu>1\) and convection is sufficiently supercritical. Moreover, 
the critical Rayleigh number varies as \(Ra_c\sim E^{-4/3}\) \citep{busse70}, 
and we finally find \(Ra/Ra_c\sim E^{1/3}\) for models at constant Nusselt 
number. For \(Nu\approx 2.2\), \(3.5\), and \(7\), we considered a maximum
Rayleigh number of \(6\), \(15\), and \(50\) times its critical 
value.

The Rossby number scaling from \cite{christensen06} together with the Nusselt 
number scaling yields
\begin{equation}
Ro\sim ((Nu-1)\,E/Pr)^{0.77}.
\label{ap:4}
\end{equation}
With \(Rm=Ro\,Pm/E\) we conclude that the magnetic Reynolds number increases 
slightly with increasing rotation rate, \(Rm\sim E^{-0.23}\). 
For \(Nu\approx 2.2\) and \(Pm=4\), for instance, we obtained \(Rm\approx130\) 
at \(E=10^{-3}\) and \(Rm\approx 270\) at \(E=10^{-5}\).  
\label{lastpage}
\end{document}